# AGE-INDEPENDENT PROCESS IN AGING HARD-SPHERE SUSPENSIONS: THE BETA PROCESS AND ITS RAMIFICATIONS.

W. van Megen

Department of Applied Physics, Royal Melbourne Institute of Technology, Melbourne, Victoria 3000, Australia.

H. J. Schöpe

Institut für Angewandte Physik, Universität Tübingen, Auf der Morgenstelle 10, 72076 Tübingen, Germany

ABSTRACT

We consider the dynamics of a suspension of hard sphere-like particles in the proximity of its glass transition, the region where the intermediate scattering functions show significant aging. The time correlation function of the longitudinal particle current shows no dependence on age and reveals behaviour of ideal super-packed fluid and glass. The power laws of the beta process of the idealised mode coupling theory are exposed directly without reliance on fitting parameters. We proffer a mechanism linking the reversible/ageless dynamics, which constitutes the beta-process, and the irreversible aging dynamics. The latter verifies predictions of spin-glass theories.

When a liquid is cooled below its freezing point quickly enough to bypass crystallisation its structural relaxation time and resistance to flow increase sharply and can do so to such an extent that, with sufficient (under) cooling, the liquid vitrifies. The physics of this phenomenon continues to be one of the more intriguing and studied subjects of classical condensed matter [1-5]. While numerous models and theories proffer explanations it's fair to say that much of the interest witnessed over recent decades, particularly that concerned with structural glasses, has been motivated by mode coupling theory (MCT) [6]. Its basic and commonly applied version predicts a sharp transition from fluid to amorphous solid – ideal glass, where a fraction of the structure is permanently arrested at a critical temperature or, in the case of a system of hard spheres considered here, a critical packing fraction, $\phi_c$. Quintessentially, the theory, to leading order in the separation from this critical point, then predicts that the decay of the time correlation function of the particle number density toward the arrested structure is governed by a power law, which, in the super-packed fluid, crosses over to a second power law. The exponents of the power laws derive from the inter-particle potential and are independent of the spatial frequency. Ostensibly, these processes, collectively referred to as the β- process, characterise rattling of particles in an arrested amorphous structure; particles in their respective neighbour cages. Final relaxation of the fluid's structure, effected through exchanges of caged particles, is governed by a master function that is the α-process. These detailed predictions have been uncovered in various computer simulations and experiments [6, 7]. In this regard MCT is without peer and, quite possibly, for this reason it has been subjected to considerable scrutiny and critique [3, 8-11]. The following are among the more contentious issues;

1. When real materials are supercooled irreversible, aging processes cut off the sharp transition from fluid to glass; as such MCT can apply only to the moderately supercooled state where irreversible processes are negligible. Even for colloidal systems, where the suspending liquid supposedly damps energy exchanges among the particles, aging is evident in the vicinity of the glass transition (GT) [12-16]. In light of this more recent work the initial notion [6, 17, 18] that irreversible dynamics are absent from the colloidal GT needs reconsideration.

2. Analyses of the time correlation functions obtained by computer simulations on simple atomic fluids [19, 20] and dynamic light scattering (DLS) experiments on suspensions of hard spheres [17, 21] in terms of MCT require several fitting parameters which are accorded considerable flexibility due to the said rounding of the transition; experimental and computational support for MCT is not un qualified.

More fundamentally, as shown by molecular dynamics (MD) and Brownian dynamics (BD) computer simulations, MCT predictions independent of the microscopic dynamics [22, 23]. So the question is, what is the nature of the seemingly slower collective dynamics consequent on excluded volume effects that emerge when the packing fraction approaches $\phi_c$ and, furthermore, what can be learned about these processes by (re-) consideration of the above two issues?

In this Letter we aim to address these issues with the support of data obtained from previous dynamic light scattering (DLS) experiments on suspensions of hard sphere like polymer particles [12, 24]. These have average radius, R=185nm, and polydispersity of approximately 7%. The latter is sufficient to delay the onset of nucleation long enough to allow stationary time correlation functions of the super-packed suspension to be measured [25, 26]. Other experimental details are documented elsewhere [12, 27, 28]. The only control parameter is the packing fraction, $\phi$, the relevant values of which locate the freezing point at that of the ideal hard sphere system, $\phi_f$=0.494 [29, 30], and a glass transition at $\phi_g \approx 0.570$ [17, 31]. In the results presented below the spatial frequency, q, and all lengths are expressed in terms of R and times in units of the Brownian time, $R^2/(6D_0)$=0.015s, where $D_0$ is the free particle diffusion constant.

We begin with the intermediate scattering function (ISF)

$$F(q,\tau;t_w) = \langle \rho(q,t_w)\rho^*(q,t_w+\tau) \rangle / \langle |\rho(q)|^2 \rangle, \qquad (1)$$

where $\rho(q,t) = \sum_{k=1}^{N} \exp[-i\mathbf{q}\cdot\mathbf{r}_k(t)]$ is the $q^{th}$ spatial Fourier component of the particle number density, $\mathbf{r}_k(t)$ the position of the $k^{th}$ particle at time t, and * indicates complex conjugation. Typical results are shown in Fig. 1. One sees that dependence on waiting time, $t_w$, is not evident for $\phi$=0.555, marginal for $\phi$=0.563 but pronounced for $\phi$=0.58. In all cases the decays of $F(q,\tau;t_w)$ from their respective plateaux, f(q), can be approximated by the stretched exponential (SE),

$$F(q,\tau>\tau_m,t_w) = f(q)\exp\left[-(\tau/\tau_s)^\beta\right]. \qquad (2)$$

$\tau_m$ is the delay time where $F(q,\tau;t_w)$ has a change in curvature, or plateau $f(q)=F(q,\tau_m;t_w)$. Values of stretching exponents of $F(q,\tau>\tau_m;t_w)$ (Eq. (2)) are independent of $t_w$ and read $\beta\approx 0.9$, $\beta\approx 0.5$, and $\beta\approx 0.9$, respectively for the three packing fractions, 0.555, 0.563 and 0.580 shown.

For $\phi=0.555$ $F(q,\tau>\tau_m;t_w)$ decays almost to zero. But for the two other cases shown the decay of $F(q,\tau>\tau_m;t_w)$ is limited by the (practical) experimental time limit. Within this limit and attendant proximity to the plateau the SE can be approximated as $\exp[-(\tau/\tau_s)^\beta]\approx 1-(\tau/\tau_s)^\beta$. Further, the inset in Fig. 1b shows, for $\phi=0.580$, that the decays $F(q,\tau>\tau_m;t_w)$ superpose when the delay time is scaled by $\tau_s$; $\tau_s$ is the only parameter in the SE that depends on $t_w$. For the colloidal glasses generally[12], $\tau_s$ increases approximately linearly with $t_w$ up to a time (approximately 120 hr.) after which the difference of $F(q,\tau>\tau_m;t_w)$ from $f(q)$ can no longer be discerned from experimental noise. Colloidal fluids show no dependence on $t_w$ except, as illustrated in Fig. 1a for $\phi=0.563$, just below the GT where $F(q,\tau;t_w)$ saturates after approximately 10 hrs.

The corresponding time correlation functions [32, 33]

$$C(q,\tau,t_w)=q^2\langle j(q,t_w)j^*(q,t_w+\tau)\rangle=-d^2F(q,\tau;t_w)/d\tau^2. \qquad (3)$$

of the longitudinal current $j(q,t) = \sum_{k=1}^{N} \hat{\mathbf{q}}\cdot\mathbf{v}_k(t)\exp[-i\mathbf{q}\cdot\mathbf{r}_k(t)]$, where $\mathbf{v}_k(t)$ is the velocity of particle k at time t, are obtained by numerical differentiation of $F(q,\tau;t_w)$. Absolute values of $C(q,\tau;t_w)$ in Fig. 2 are shown because it is negative. So $C(q,\tau;t_w)$ exposes correlation of particle current *reversals* directed along the propagation vector, $\mathbf{q}$; deviation of $C(q,\tau;t_w)$ from an exponentially decaying function of $\tau$ exposes persistence of anisotropy in the particles' motion. As foreshadowed in Ref. [24] for the colloidal glass, we see here more generally that any dependence on waiting time evident in $F(q,\tau;t_w)$ appears not to have been transferred to $C(q,\tau)$. We proffer an explanation below and for now regard this result a fortuitous outcome that exposes, in $C(q,\tau)$, the dynamics around the GT free of aging.

In the colloidal glass (Fig. 2b) the initial decay $C(q,\tau)$, follows a SE which, being an accumulation of overdamped current reversals, exposes remnant memory of the particles' Brownian motion. In addition we see here a more persistent, non-Brownian process whose emergence is exposed by a crossover, at $\tau=\tau_c\approx 1$, from the SE to another decay that can be described most simply by a power law; $C(q,\tau>\tau_c)=A\tau^{-\nu}$. The results of the latter fitted to $C(q,\tau)$ from $\tau=1$ to the experimental noise floor, $\tau_\infty\approx 10^6$, for several values of $\phi>\phi_g$ and q are shown in Fig. 3. The exponent $\nu=2.27$ ($\pm 0.07$) shows no systematic variation with q. Accordingly we express the amplitude as $A=A_0 h(q)$ and adjust $A_0$ so that $h(q\approx q_m)=1$. Here $q_m$ is the position of the primary maximum of the structure factor. Despite considerable errors, values of $h(q)$ systematically have minima around $q_m$ where also the amplitude, $f(q)$, of the arrested structure is largest [34]. Within these experimental and numerical errors $A_0$ shows no systematic variation with $\phi$.

For the two colloidal fluids (Fig. 2a) $C(q,\tau)$ again crosses over at $\tau=\tau_c\approx 1$ from a stretched exponential to a power law but then, for $\phi=0.563$, there appears another cross over, at $\tau=\tau_m\approx 10^3$, to a second power law. The power laws are merely the simplest, albeit not necessarily unique, approximations to the decay of $C(q,\tau)$ between $\tau_c$ and $\tau_m$ and between $\tau_m$ and the noise floor. Despite larger errors, due to the more limited fitting ranges in this case, the values of respective exponents, $2.30\pm 0.08$ and $1.52\pm 0.04$ (Fig. 3a), show no systematic variation with q. In addition, the amplitudes of both power laws (Fig. 3b) have minima around $q_m$. A similar cross over and delineation of two power laws is not evident for the slightly lower packing fraction, $\phi=0.555$.

According to MCT the decay of the ISF to/from the plateau is [6, 7]

$$F_{MCT}(q, t_0 \ll \tau \ll \tau_\alpha) = f(q) + |\sigma|^{1/2} h(q) g_\pm(\tau/\tau_\beta). \tag{4}$$

Here $t_o$ and $\tau_\alpha$ are the time scales that characterise the microscopic motion and the final $\alpha$-relaxation of the structure of the super-packed fluid, $\sigma = c_0(\phi - \phi_c)\phi_c$ and $\tau_\beta = t_0|\sigma|^{-1/2a}$. The decay to $f(q)$ is described by the first power law, $g_+(\tau < \tau_m) = \tau^{-a}$. In the fluid this crosses over to the second power law $g_-(\tau) = -B\tau^b$. For the hard sphere interaction values of the exponents are a=0.30 and b=0.54 [35]. The current correlator derived from Eq. (4) can be written as

$$C_{MCT}(q, \tau) = A_0 h(q) c_\pm(\tau), \tag{5}$$

where the values of the exponents of the power laws, $c_+(\tau) \sim \tau^{a'}$ and $c_-(\tau) \sim \tau^{b'}$, are now $a' = -a-2 = -2.30$ and $b' = b-2 = -1.46$. These and the critical amplitudes, $h(q)$, from the MCT are shown in Fig. 3 along with the experimental results.

Thus, in the vicinity of the GT, $\phi \gtrsim 0.56$ in this case, we find the following predictions of the MCT *without resorting to adjustable parameters*; First, the power laws and the values of their respective exponents are consistent with those of the theory. Second, the space-time factorisation of the $\beta$ process is verified insofar that these exponents show no systematic variation with q. Third, the amplitudes of the power laws consistently show minima at spatial frequency, $q_m$, where the amplitude of the arrested structure is largest. To place these results in perspective we point out that in previous tests of MCT based on similar DLS data [17], though of lesser quality, the functions $g_\pm(\tau/\tau_\beta)$ were taken as given by the theory and fitted to $F(q,\tau)$ using $\sigma$, $t_0$ and $f(q)$ as adjustable parameters.

The algebraic decay of the ISF to the plateau (Fig. 1) expresses at once the time scale invariance indicative of an intermittent process and, by its transfer to the current correlator (Fig. 2), a process that's anisotropic. Furthermore, the consistency of the DLS results, along with those of molecular *and* Brownian dynamics computer simulations [22, 23, 36], with MCT implies the dynamics expressed by the $\beta$-process lacks memory of the microscopic kinetics. In the theory the latter are differentiated only by the time scale $t_0$. The implication for the colloidal glass is that the close proximity of the particles mitigates the dissipative efficacy of the solvent. As a result (see Fig. 2) the correlation of overdamped current reversals is survived, beyond the delay time $\tau_c$, by systematic displacement reversals; a situation redolent of "collisions" in an amorphous assembly of "rattling" particles.

Further insight may be gained by considering the plateau values, $R_m = \sqrt{\langle \Delta r(\tau_m)^2 \rangle}$ of the root mean squared displacement (RMSD). Results derived from experiment and theory [37] are shown in Fig. 4. The ratios, $R_m/R_c$, where $R_c = [\phi_R/\phi]^{1/3} - 1$ is the average distance between particle surfaces and $\phi_R = 0.64$ is the random close packing value, are also shown. Errors notwithstanding, we see that experiment and theory are not inconsistent. The overall monotonic decrease in $R_m$ with $\phi$ might be seen as indicative of increasing localisation of particles in their temporary ($\phi < \phi_g$) or permanent ($\phi > \phi_g$) neighbour cages. However, the ratio, $R_m/R_c$, suggests a cooperation among cage fluctuations that allows particles to move, on average, some 5-8 times the inter-surface distance. The picture here is reminiscent of intermittent jumps involving of some 10 or so particles in strings revealed by numerous microscopic stdies[38-44]. The current correlator, $C(q, \tau > \tau_c)$, exposes the directionality/anisotropy of this process more explicitly than correlators, such as the ISF, of scalar variables.

What is not exposed by $C(q,\tau>\tau_c)$, having been effectively subtracted from the experimental data by taking the second derivative of $F(q,\tau)$ and explicitly excluded from the MCT, are those jumps, or their accumulations that are irreversible; Aging in other words. (See Ref.[44] for an illustration of the distinction between the two.) We see from Fig. 1 that the contribution of aging to the ISF becomes significant for $\tau>\tau_m$. Note, since the non-ergodicity parameter, $f(q)$, is independent of waiting time, reversible and irreversible jumps both express collective dynamics of the one rattling amorphous system. So aging can only be and is seen to emerge here as an additive, but irreversible, continuation of the β-process. In these respects and in that they satisfy waiting time superposition (Fig. 1b, inset) these results support the predictions of mean field spin glass theories [45-47].

Further explanation of the above opens with scenario inferred from several previous experiments [48-51] that caging features only in the super-packed ($\phi_f<\phi<\phi_f$) suspension but not in the thermodynamically stable suspension ($\phi<\phi_f$). In the latter $C(q,\tau)$ follows a SE to the noise floor which delay/stretching is effected by the correlation of over-damped/diffusing current reversals [51]. Then as $\phi$ exceeds $\phi_f$ the emergence of caging is inferred from the occurrence of a crossover at $\tau_c$ to another source of delay in the decay of $C(q,\tau)$ around $q_m$. With increasing $\phi$ caging spreads in a widening window, $q_m\pm\delta q$, of spatial frequencies centred around $q_m$. Concomitantly, the frequency of cage exchanges decreases in approximate proportion to the long-time diffusion constant [49, 51, 52]. In this way $C(q\notin[q_m\pm\delta q],\tau)$ and $C(q\in[q_m\pm\delta q],\tau>\tau_c)$ express the decay of diffusing and caged "modes"; their anisotropy is understood since we are still considering, in the current correlator, just the components of the particles' motions in the direction of the propagation vector.

The main feature of the diffusing modes, whether in the thermodynamically stable or super-packed suspension, is that all dependence on q resides in the short-time diffusion coefficient, $D(q)$, as obtained from the initial decay of the ISF, $F(q,\tau\ll\tau_m)=\exp[-D(q)q^2\tau]$ [53]. So when the delay time is expressed as $\tau^*=D(q)q^2\tau$ the resulting scaled current correlators $C^*(\forall q,\tau^*)$, for $\phi<\phi_f$, and $C^*(q\notin[q_m\pm\delta q],\tau^*)$, for ($\phi_f<\phi<\phi_g$), are independent of $q$[49, 51]. Moreover, the decay time, $\tau_x^*$, of the SE fitted to $C^*(q,\tau^*)$ is less than one and decreases with $\phi$. So, as discussed in detail in Ref. [51]35, the collective, *anisotropic* diffusing modes, $C^*(q,\tau^*)$, explore configuration space more efficiently than the underpinning, *isotropic* Brownian fluctuations.

Together, diffusing and caged modes presents another perspective of the notion of coexisting dynamical phases, respectively expressing fluidity and solidity, as well as the reciprocal space complement to the more commonly adopted direct space expression of dynamical heterogeneity [3, 10, 39, 54-58].

With the above scenario we recognise some of the rudiments of MCT;

(i) A β-process by reversible jumps in the caged modes in, and confined to, the window of spatial frequencies $q\in[q_m\pm\delta q]$;

(ii) An α-process by exchanges of caged particles achieved through coupling of the caged modes. Such exchanges can be mediated only by the diffusing modes in the complementary window, $q \notin [q_m \pm \delta q]$. However, such exchanges become rarer when, on increasing $\phi$, the window of caged modes expands and, concomitantly, the complementary window of diffusing modes shrinks. Note, since caging is implicit in the MCT, it does not recognise its onset at $\phi_f$ as found for the actual hard sphere system [17, 51, 59].

(iii) (Definition of) a critical super-packing, $\phi_c$, approached as $\delta q \rightarrow q_m$ [51] and large scale diffusion is squeezed out by the caged modes. In this limit we attain the ideal glass in which diffusing modes exist only for $q > 2q_m$ or for wavelengths less than $\pi/q_m$. In this case there's no mechanism for particles to exchange position, at least not by diffusive motion. When, for $\tau > \tau_c$, these short wavelength diffusing modes have relaxed the $\beta$-process that characterises the (reversible) dynamics of the rattling amorphous solid is exposed.

(iv) The asymptotic results of MCT (Eq. (4), (5)) are recovered as $\delta q \rightarrow q_m^-$ ($\phi \rightarrow \phi_c^-$) and the dynamics are dominated by the rattling amorphous structure, manifested by the first power law of the $\beta$-process. At the same time the second power law is exposed as a result of the increasing rarity/intermittency of exchanges of caged particles.

The packing fraction, $\phi=0.563$, is evidently close enough to $\phi_c$ ($=\phi_g \approx 0.570$) to expose both power laws identified in $F(q,\tau>\tau_m)$ (Fig. 1a) and $C(q,\tau>\tau_m)$ (Fig. 2a). From these same Figures it is also evident that lowering $\phi$ by little more that 1%, from 0.563 to 0.555, is sufficient for the second power law, $F(q,\tau>\tau_m) \sim -\tau^{0.5}$, of the $\beta$ process to be replaced by a SE, $F(q,\tau>\tau_m) \sim \exp[-\tau^{0.9}]$, predicted for the $\alpha$ process [6, 60]. This is effectively removed in derivation of the corresponding current correlator with the result that $C(q,\tau_x<\tau<\tau_\infty)$, can still be approximated by a power law $\sim \tau^{a'}$ (Fig. 2a). However, we cannot identify the latter with the first power law of the $\beta$ process because the exponent now shows a systematic variation with q that cannot be accommodated by experimental noise (Fig. 3a).

As perhaps first pointed out by Rahman [61] and evident from the numerous more recent microscopic studies already mentioned, particles cooperate in their movement by following each other. The anisotropy of this mechanism is explicit in the current correlators of diffusing and caged modes, $C^*(q \notin [q_m \pm \delta q], \tau^*)$ and $C(q \in [q_m \pm \delta q], \tau)$. As mentioned above, by their independence of q the diffusing modes are not coupled to the structure. But caged modes are only weakly coupled to it. This is evident, for example, from the fact that the (first) peak to trough ratio of $h(q)$ or $f(q)$ is less than 20% of that of $S(q)$ [17, 34, 62]. This weak, or lack of, coupling to the structure underpins the efficiency by which anisotropic, collective particle motions explore configuration space. This has further consequences; First and foremost, the intermittent, jumps, reversible and irreversible are only weakly correlated. So, their continuum manifestation, that emerges with increasing delay time is expressed in the ISF by a weakly stretched exponential $F(q,\tau>\tau_m;t_w) \sim \exp[-\tau^\beta] \approx 1-\tau^\beta$. The value of the exponent ($\beta \approx 0.9$) is evidently close enough to one that the decay of the ISF from the plateau in the experimental window approximates closely enough to a linear function of $\tau$ for its second time derivative (Eq. (3)) to be absorbed by experimental noise. As a result the current correlator appears ageless. Second, aging can be approximated by continuous time random walks [41]. Third, as MCT demonstrates, the phenomenological consequences of caging can be captured with the simplest (quadratic) coupling of caged modes.

Before concluding we digress here to consider how scenarios other than the above aging can result from different particle size distributions (PSD). A much narrower PSD than that used in the present work would enhance nucleation of the crystal phase and preclude study of the deeply super-packed state [25, 63]. Differential arrest is another scenario; where an initial arrested structure formed by the bulk of near-average sized particles is slowly relaxed (melted) by a sub-population of small mobile particles. The resulting slow, but still reversible structural fluctuations effect a waiting time dependent but otherwise complete decay of the ISF. The mechanism was first exposed in DLS experiments of binary mixtures [64] and more recently in simulations and further experiments [65-67]. Clearly this mechanism is sensitive to the presence of a population of small particles in the PSD be that in a multi-component mixture, a broad but continuous symmetrical PSD or one that's negatively skewed. Any could incur differential arrest. The PSDs of the PMMA based particles used so widely as model hard sphere colloids tend to be negatively skewed [28, 63]. Moreover, probably due to varying degrees of secondary nucleation different preparations of particles having similar average radii and polydispersities may have quite different skewnesses. So, in one case we see the GT scenario, studied above, where an age-dependent decay of the ISF saturates to an arrested structure; $F(q,\tau \rightarrow \infty; t_w \rightarrow \infty) = f_c(q)$. In another, due to differential arrest, decay of the ISF is age-dependent but complete; $F(q, \tau \rightarrow \infty; t_w) = 0$ [15, 16]. These different scenarios have aroused controversy about the dynamics around the GT[65, 67-71]. However, we suggest they are just different but interesting rather than conflicting scenarios.

*Conclusion.* By isolating, via the current correlator, the stationary dynamics in an otherwise aging super-packed suspension of hard-sphere particles allows one to define a sharp transition to an ideal glass. We identify the collective mechanism that underpins the algebraic decays of the ISF to/from its arrested structure and, thereby the β-process of mode coupling theory. At the same time the more salient predictions of the idealised version of the theory are quantitatively exposed without recourse to the theory as such. What's more, the perspective of the dynamics of the ideal GT presented also gives insight into non-stationary/aging processes that round the ideal GT.

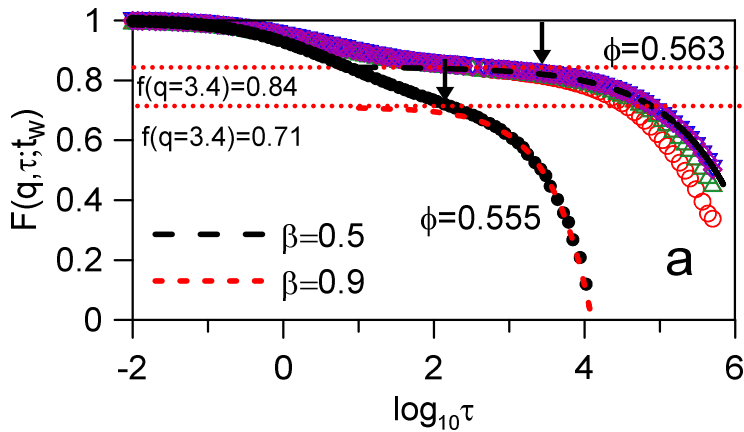

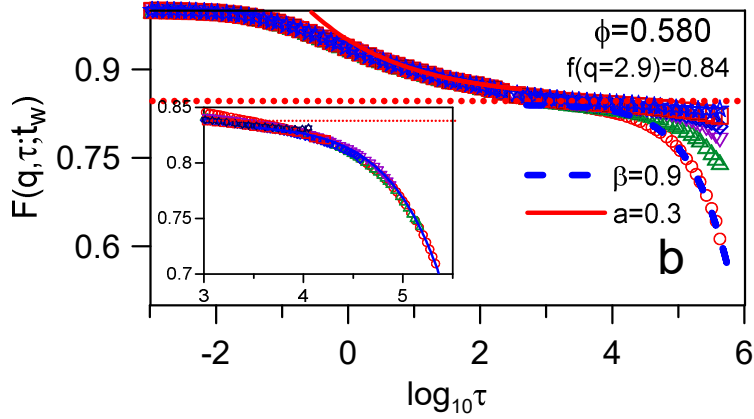

Fig. 1. Intermediate scattering functions, $F(q,\tau)$, for indicated packing fractions, $\phi$, spatial frequencies, $q$, plateaux or non-ergodicity parameters, $f(q)$ (red dotted lines), and waiting times $t_w$=1h (red circle,○), 4h (green triangle,△), 10h (purple inverted triangle, ▽), 24h (blue diamond, ◇ ), 54h (red square, □ ), 120h (blue star,☆ ). Various dashed lines that follow the decay from $f(q)$ are SEs, $f(q)\exp[-(\tau/\tau_s)^\beta]$ for values of $\beta$ shown. Solid line is a power law with indicated exponent. Downward pointing arrows indicate location of cross over times $\tau_m$. The insert in (b) shows the decays, $F(q,\tau>\tau_m;t_w)$ versus $\log_{10}(\tau/\tau_s)$. Note differences in axes' scales of (a), (b) and inset.

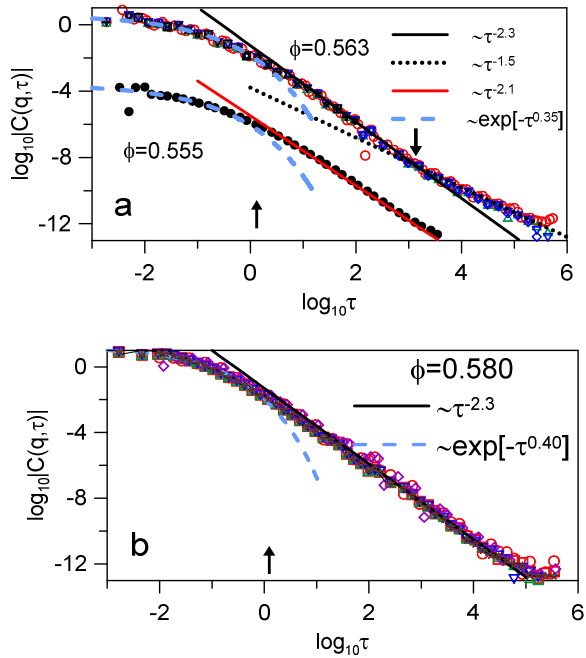

Fig. 2. Current correlators, $C(q,\tau)$, derived from ISFs, $F(q,\tau)$ in Fig. 1, for packing fractions indicated. Different (coloured) symbols are for values of $t_w$ listed in Fig. 1. Upward and downward pointing arrows indicate location of cross over times $\tau_x$ and $\tau_m$ defined in the text. Various stretched exponential and power law functions and their respective exponents are also shown.

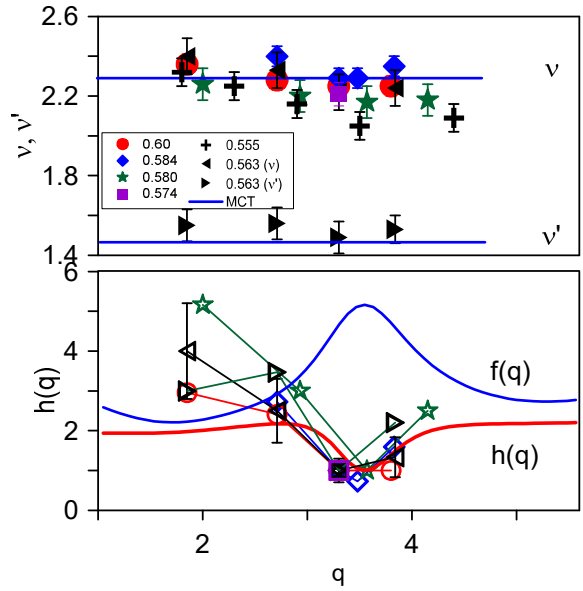

Fig. 3. Top panel; Experimental exponents, $\nu$ and $\nu'$, of the first and second power laws obtained from the current correlator for values of $\phi$ indicated. Solid lines are the corresponding MCT exponents, a' and b'. Bottom panel. Experimental (symbols) and MCT (red line) critical amplitudes, $h(q)$. $f(q)$ is the MCT result for the fraction of the arrested structure.

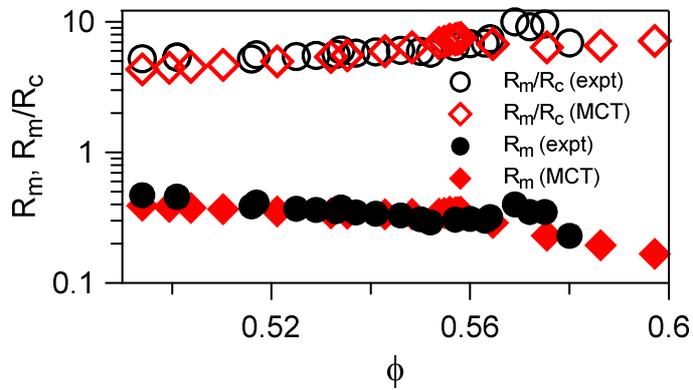

Fig. 4.   RMSD values.

ACKNOWLEGEMENTS

The authors acknowledge contributions to this work from Gary Bryant, Juergen Horbach, Vincent Martinez and Thomas Voigtmann.

REFERENCES

1.	F. Sciortino and P. Tartaglia, Advances in Physics **54** (6-7), 471-524 (2005).
2.	A. Cavagna, Physics Reports-Review Section of Physics Letters **476** (4-6), 51-124 (2009).
3.	K. Binder and W. Kob, *Glassy materials and disordered solids; An introduction to their statistical mechanics*. (World Scientific, london, 2011).
4.	S. P. Das, *Liquids at Freezing and Beyond*. (Cambridge University Press, Cambridge, 2011).
5.	L. Berthier and T. A. Witten, Physical Review E **80** (2), 15 (2009).
6.	W. Götze, *Complex Dynamics of Glass-Forming Liquids; A Mode-Coupling Theory*. (Oxford University Press, Oxford, 2009).
7.	W. Götze, J. Phys.-Condes. Matter **11** (10A), A1-A45 (1999).
8.	S. P. Das, Reviews of Modern Physics **76** (3), 785-851 (2004).
9.	D. R. Reichman and P. Charbonneau, J. Stat. Mech.-Theory Exp., 23 (2005).
10.	L. Berthier and G. Biroli, Reviews of Modern Physics **83** (2), 587-645 (2011).
11.	S. Gokhale, A. K. Sood and R. Ganapathy, Advances in Physics **65** (4), 363-452 (2016).
12.	V. A. Martinez, G. Bryant and W. van Megen, J. Chem. Phys. **133** (11) (2010).
13.	G. L. Hunter and E. R. Weeks, Reports on Progress in Physics **75** (6) (2012).
14.	L. Cipelletti and L. Ramos, J. Phys.-Condes. Matter **17** (6), R253-R285 (2005).
15.	G. Brambilla, D. El Masri, M. Pierno, L. Berthier, L. Cipelletti, G. Petekidis and A. B. Schofield, Phys. Rev. Lett. **102** (8), 4 (2009).
16.	D. El Masri, G. Brambilla, M. Pierno, G. Petekidis, A. B. Schofield, L. Berthier and L. Cipelletti, J. Stat. Mech.-Theory Exp., 28 (2009).
17.	W. van Megen and S. M. Underwood, Physical Review E **49** (5), 4206-4220 (1994).
18.	G. Szamel and E. Flenner, Europhysics Letters **67** (5), 779-785 (2004).
19.	W. Kob and H. C. Andersen, Physical Review E **51** (5), 4626-4641 (1995).
20.	W. Kob and H. C. Andersen, Physical Review E **52** (4), 4134-4153 (1995).
21.	E. Bartsch, V. Frenz and H. Sillescu, Journal of Non-Crystalline Solids **172**, 88-97 (1994).
22.	H. Löwen, J. P. Hansen and J. N. Roux, Physical Review A **44** (2), 1169-1181 (1991).
23.	T. Gleim, W. Kob and K. Binder, Phys. Rev. Lett. **81** (20), 4404-4407 (1998).
24.	V. A. Martinez, G. Bryant and W. van Megen, Phys. Rev. Lett. **101** (13) (2008).
25.	S. I. Henderson and W. van Megen, Phys. Rev. Lett. **80** (4), 877-880 (1998).
26.	S. I. Henderson, T. C. Mortensen, S. M. Underwood and W. vanMegen, Physica A **233** (1-2), 102-116 (1996).
27.	S. M. Underwood and W. van Megen, Colloid and Polymer Science **274** (11), 1072-1080 (1996).
28.	G. Bryant, S. Martin, A. Budi and W. van Megen, Langmuir **19** (3), 616-621 (2003).
29.	P. N. Pusey and W. van Megen, Nature **320** (6060), 340-342 (1986).
30.	S. M. Underwood, J. R. Taylor and W. van Megen, Langmuir **10** (10), 3550-3554 (1994).

31. P. N. Pusey and W. van Megen, Phys. Rev. Lett. **59** (18), 2083-2086 (1987).
32. J. P. Boon and S. Yip, *Molecular Hydrdynamics*. (Dover, New York, 1980).
33. J.-P. Hansen and I. R. McDonald, *Theory of Simple Liquids*, 4th ed. (Academic Press, AMSTERDAM, 2013).
34. W. van Megen, S. M. Underwood and P. N. Pusey, Phys. Rev. Lett. **67** (12), 1586-1589 (1991).
35. W. Götze and L. Sjögren, Physical Review A **43** (10), 5442-5448 (1991).
36. A. M. Puertas, J. Phys.-Condes. Matter **22** (10), 6 (2010).
37. W. van Megen, Physical Review E **76** (6) (2007).
38. C. Donati, J. F. Douglas, W. Kob, S. J. Plimpton, P. H. Poole and S. C. Glotzer, Phys. Rev. Lett. **80** (11), 2338-2341 (1998).
39. S. C. Glotzer, Journal of Non-Crystalline Solids **274** (1-3), 342-355 (2000).
40. T. Kawasaki and A. Onuki, Physical Review E **87** (1), 10 (2013).
41. M. P. Ciamarra, R. Pastore and A. Coniglio, Soft Matter **12** (2), 358-366 (2016).
42. W. Kob and J. L. Barrat, European Physical Journal B **13** (2), 319-333 (2000).
43. P. M. de Hijes, P. Rosales-Pelaez, C. Valeriani, P. N. Pusey and E. Sanz, Physical Review E **96** (2), 6 (2017).
44. K. Vollmayr-Lee, J. Chem. Phys. **121** (10), 4781-4794 (2004).
45. W. Kob and J. L. Barrat, Phys. Rev. Lett. **78** (24), 4581-4584 (1997).
46. J. P. Bouchaud, L. Cugliandolo, J. Kurchan and M. Mezard, Physica a-Statistical Mechanics and Its Applications **226** (3-4), 243-273 (1996).
47. J. P. Bouchaud, in *Anomalous Diffusion: From Basics to Applications*, edited by A. Pekalski and K. Sznajd-Weron (Springer-Verlag Berlin, Berlin, 1999), Vol. 519, pp. 140-150.
48. W. van Megen, V. A. Martinez and G. Bryant, Phys. Rev. Lett. **102** (16) (2009).
49. W. van Megen, V. A. Martinez and G. Bryant, Phys. Rev. Lett. **103** (25) (2009).
50. M. Franke, S. Golde and H. J. Schöpe, in *4th International Symposium on Slow Dynamics in Complex Systems: Tohoku Univ*, edited by M. Tokuyama and I. Oppenheim (2013), Vol. 1518, pp. 214-221.
51. W. van Megen and H. J. Schöpe, J. Chem. Phys. **146** (10), 9 (2017).
52. W. van Megen, T. C. Mortensen, S. R. Williams and J. Müller, Physical Review E **58** (5), 6073-6085 (1998).
53. P. N. Pusey, in *Liquids, Freezing and Glass Transition, Part II*, edited by J. P. Hansen, D. Levesque and J. Zinnjustin (North Holland, Amsterdam, 1991), Vol. 2, pp. 764-942.
54. W. K. Kegel and A. van Blaaderen, Science **287** (5451), 290-293 (2000).
55. H. Sillescu, Journal of Non-Crystalline Solids **243** (2-3), 81-108 (1999).
56. M. D. Ediger, Annual Review of Physical Chemistry **51**, 99-128 (2000).
57. L. Berthier, Biroli, G., Boachaud, J-P, Cipelletti, L. and van Saarloos, W., *Dynamical Heterogeneities in Glasses, Colloids and Granular Media*. (Oxford University Press, Oxford, 2011).
58. R. Pastore, A. Coniglio, A. de Candia, A. Fierro and M. P. Ciamarra, J. Stat. Mech.-Theory Exp. (2016).
59. V. A. Martinez, E. Zaccarelli, E. Sanz, C. Valeriani and W. van Megen, Nat. Commun. **5** (2014).
60. M. Fuchs, I. Hofacker and A. Latz, Physical Review A **45** (2), 898-912 (1992).
61. A. Rahman, J. Chem. Phys. **45** (7), 2585-& (1966).
62. W. van Megen and S. M. Underwood, Physical Review E **47** (1), 248-261 (1993).
63. H. J. Schöpe, G. Bryant and W. van Megen, J. Chem. Phys. **127** (8) (2007).
64. S. R. Williams and W. van Megen, Physical Review E **64** (4) (2001).

65. D. Heckendorf, K. J. Mutch, S. U. Egelhaaf and M. Laurati, Phys. Rev. Lett. **119** (4), 6 (2017).
66. M. Laurati, T. Sentjabrskaja, J. Ruiz-Franco, S. U. Egelhaaf and E. Zaccarelli, Phys. Chem. Chem. Phys. **20** (27), 18630-18638 (2018).
67. E. Zaccarelli, S. M. Liddle and W. C. K. Poon, Soft Matter **11** (2), 324-330 (2015).
68. W. van Megen and S. R. Williams, Phys. Rev. Lett. **104** (16) (2010).
69. G. Brambilla, D. El Masri, M. Pierno, L. Berthier, L. Cipelletti, G. Petekidis and A. Schofield, Phys. Rev. Lett. **104** (16), 1 (2010).
70. J. Reinhardt, F. Weysser and M. Fuchs, Phys. Rev. Lett. **105** (19), 1 (2010).
71. G. Brambilla, D. El Masri, M. Pierno, L. Berthier and L. Cipelletti, Phys. Rev. Lett. **105** (19), 1 (2010).